# DECT-2020 New Radio: The Next Step Towards 5G Massive Machine-Type Communications

Roman Kovalchukov, Dmitri Moltchanov, Juho Pirskanen, Joonas Säe,
Jussi Numminen, Yevgeni Koucheryavy, and Mikko Valkama

*Abstract*—Massive machine type communications (mMTC) is one of the cornerstone services that must be supported by 5G systems. 3GPP has already introduced LTE-M and NB-IoT, referred to as cellular IoT (CIoT), in 3GPP Releases 13, 14, and 15 and submitted these technologies as part of 3GPP IMT-2020 technology submission to ITU-R. Even though NB-IoT and LTE-M have been shown to satisfy 5G mMTC requirements, it is expected that these CIoT solutions will not address all aspects of IoT and ongoing digitalization, including the support for direct communication between "things" with flexible deployments, different business models, as well as support for even higher node densities and enhanced coverage. In this paper, we first describe the Digital Enhanced Cordless Telecommunications (DECT) DECT-2020 standard recently published by ETSI for mMTC communications. We then evaluate its performance and compare it to the existing low power wide area network (LPWAN) solutions showing that it outperforms those in terms of supported density of nodes while keeping delay and loss guarantees at the required level.

## I. INTRODUCTION

For many years, the Internet of Things (IoT) has been predicted to have almost exponential growth [1]. However, this progress has taken much longer than expected. With the rise of big data and artificial intelligence, IoT has become a new megatrend foreseen to impact many industry areas [2]. This progress depends on the availability of off-the-shelf enabling radio technologies.

Conventionally, applications act as a driving factor towards developing enabling technologies. Industry 4.0 concept dictates small delays in the air interface, enabling real-time synchronization and reconfiguration capabilities [3]. On the other hand, condition monitoring and logistics applications are characterized by high node density and do not pose strict requirements on delay and single packet loss performance [4]. These new service requirements are taken into account in the International Telecommunication Union Radiocommunication Sector (ITU-R) International Mobile Telecommunications-2020 (IMT-2020) process such that any IMT-2020 technology proposal shall meet all the requirements of usage scenarios to be recognized as 5G technology by ITU-R [5].

Addressing these requirements, 3rd Generation Partnership Project (3GPP) has specified two services and associated enablers – Narrowband IoT (NB-IoT), Long-Term Evolution MTC (LTE-M), and the upcoming New Radio (NR) based ultra-reliable low-latency communications (URLLC) [6]. Nevertheless, numerous upcoming applications fall between these two extremes [7], e.g., presence monitoring and remote light control. These applications pose much stricter requirements on delay and loss performance compared to massive machine type communications (mMTC). However, they are still more relaxed than those of URLLC. Furthermore, reliability and availability for these applications are becoming much more critical compared to power consumption.

Perceiving these needs, the European Telecommunications Standards Institute Digital Enhanced Cordless Telecommunications (ETSI DECT) working group initiated new efforts in early 2018 aiming to standardize a technology that would fulfill both the mMTC and URLLC requirements of the 5G and support local deployments without separate network infrastructure, network planning, and spectrum licensing agreements. Most of the ongoing efforts are concentrated on enabling advanced networking capabilities, enhancing the conventional centralized operation and star-based topology by introducing multi-hop operation and random access schemes for capacity and coverage extension. These efforts commenced in July 2020, providing a cost-efficient industry-wide mMTC enabler by ETSI DECT-2020.

The goal of this paper is to describe the technical solutions of the DECT-2020 standard and to evaluate its performance. To this aim, we start by reviewing the requirements of the new applications of mMTC technologies as well as design approaches to low power wide area network (LPWAN) technologies in Section II. We then provide the details of the recently standardized DECT-2020 mMTC solution in Section III. In Section IV, we study the proposed solution using standardized ETSI/ITU-R methodology [8] and compare it with modern LPWAN technologies. Conclusions are drawn in the last section.

## II. 5G mMTC State-of-the-Art and Requirements

### A. 5G mMTC Service Requirements

The requirements for 5G mMTC vary and depend on the application. For conventional monitoring applications, ITU-R mandates the network to achieve the node density of at least one million devices per square kilometer with the minimum message intensity of one packet per two hours [5]. At this traffic intensity, the packet loss ratio (PLR) must be at most one percent, while the maximum delay should be upper bounded by 10 seconds.

The evolving IoT market is accompanied by several trends affecting the abovementioned values. First, as the number of sensors in a single device increases, the amount of traffic generated by the device grows. In this multipurpose equipment, low power consumption becomes essential to support long operating times with coin cell battery sizes. The market is further expanded by novel services requiring much tighter delay guarantees closer to real-time requirements, such as a lighting control use case, where indoor or outdoor lights are controlled by radio communications. Additionally, in many

R. Kovalchukov, D. Moltchanov, J. Säe, Y. Koucheryavy, and M. Valkama are with Tampere University, Finland. Email: firstname.lastname@tuni.fi

J. Pirskanen and J. Numminen are with Wirepas Oy, Tampere, Finland. Email: firstname.lastname@wirepas.com

This work was supported by the Business Finland 5G-FORCE project.

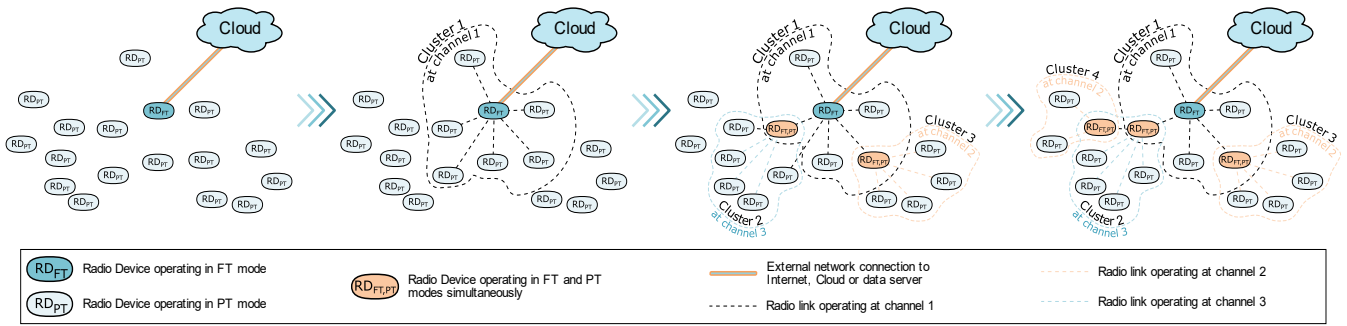

Fig. 1. Illustration of topology formation and RDs' operating modes in DECT-2020 technology.

IoT applications such as warehouse inventory or factory automation, the data communication is local, calling for a local area network deployment rather than wide-area type systems. In addition to higher traffic demands and stricter delay requirements, the reliability of the communication becomes essential. Within the scope of the light control example, should the desired operation malfunction, this can result in a non-optimal end-user experience and may lead to a significant risk of injury. Similarly, the communication must avoid a single point of failure, and operation should be possible even if the connection to the infrastructure is lost. Systems must ensure that a single communication failure does not prevent other parts from operating and that the problem is isolated to only the impacted node.

In summary, the new mMTC applications require stricter performance targets than declared in [5]. The DECT-2020 mMTC standard aims to fill this void while supporting the wide-area IoT requirements defined by ITU-R.

*B. Design Approaches and Modern Enablers*

Within the 5G mMTC technology standardization, two independent tracks are identified. The first track is based on self-configured and infrastructure-less wireless sensor networks that operate in unlicensed spectrum. This type of operation is achieved by mesh network topology, where devices can route other nodes' data to an access point or a sink with connectivity to the outside world. The second approach is utilizing a star topology network either in unlicensed/licensed spectrum, which is referred to as cellular IoT (CIoT). All nodes in these networks connect directly to the infrastructure.

The mesh-based sensor networks are typically based on Institute of Electrical and Electronics Engineers (IEEE) 802.15 solutions, such as Bluetooth, Thread, ZigBee, Wirepas Mesh [9]. The utilized radio technologies are often simple and have limited capabilities regarding the data rate, modulation schemes, and channel coding. The most significant benefit is that there is no need for network infrastructure, implying that anyone can deploy nodes without contract or subscription with a commercial or private network operator.

The star topology networks operating in unlicensed bands include Sigfox, LoRA, and Wi-Fi HaLow [10]. Sigfox is a proprietary solution, whereas LoRA is partly open for the alliance members. The CIoT approach and Wi-Fi Halow are based on open standards defined by 3GPP and IEEE. Within the 3GPP, there exist a few different technologies [10]. The first one is Extended coverage Global System for Mobile Communications IoT (EC-GSM-IoT), which is backward compatible with 2G GSM. The second supported 3GPP standard is LTE-M, a solution compatible with LTE with a lowered operating bandwidth of 1.4 MHz and reduced device complexity. The last 3GPP option is NB-IoT that operates over 200 kHz with a completely newly designed radio access.

In the star topology, device-to-device (D2D) communication is not supported. What drives the system design to maintain large link budget values to reduce the possibility of being out of coverage. Consequently, the available data rates in Sigfox, LoRA, and NB-IoT are limited due to the signal and bandwidth design.

To support high node densities and satisfy stricter delay and loss guarantees, network designers may consider expanding the frequency resources available for mMTC communications, resulting in increased capital expenditures for mMTC operators. Alternatively, one may rely on spatial frequency reuse by employing multi-hop relaying. The use of mesh systems induces its own shortcomings, including the need for routing functionality at end systems and complex random access protocols. However, they can be alleviated by introducing special topologies such as trees or directed acyclic graphs that do not require explicit routing. Thus, several telecommunications actors consider multi-hop systems as a potential solution for 5G.

III. DECT-2020 mMTC Standard

*A. System Architecture*

The system architecture of DECT-2020, described in multi-part ETSI standard [11], is cluster-tree mesh network topology without dedicated network infrastructure nodes, as shown in Fig. 1. The first part of the standard specifies radio devices (RD) that may operate in fixed (FT) or portable termination (PT) modes or both modes simultaneously. In the FT mode, an RD applies a radio coordination function to manage local radio resources and enable other RDs in PT mode to connect to it. The RD, operating in FT mode, routes data from these RDs and its own data in PT mode towards the next hop RD operating in FT mode, reaching the RD in FT mode with an external internet connection. Thus, the same RD can also operate in FT mode to provide connectivity to another set of RDs, as shown in Fig. 1.

At the network initiation, only RDs with direct connection to the external network operate in FT mode. Usually, these RDs are sink nodes, which are neither traffic sources nor destinations. Once RD in PT mode establishes a connection to an RD in FT mode, it can also start operating in FT mode to provide connection to other RDs. When alternative connection is available for RD in PT mode, it can switch between RDs in FT mode depending on the configuration.

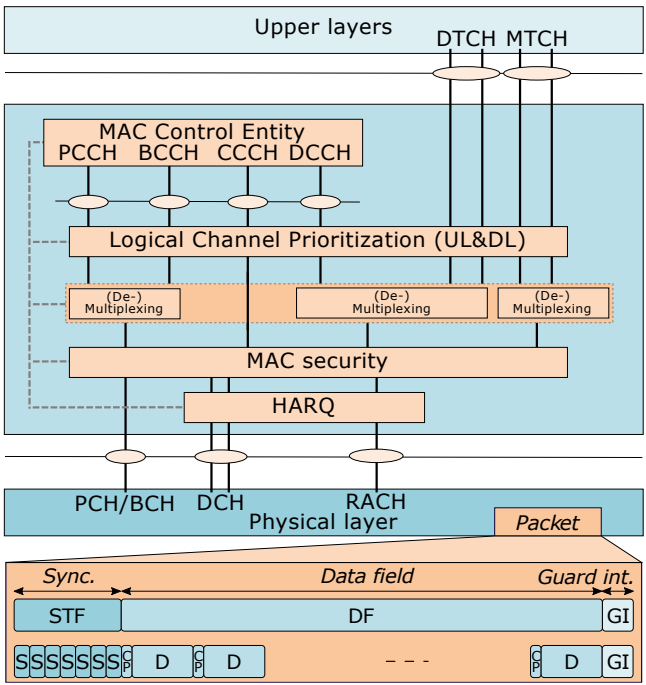

Fig. 2. Protocol stack architecture and packet structure in DECT-2020.

For high scalability, the devices are free to choose the operation mode locally. The external internet connectivity is outside of DECT-2020 standard, allowing the use of different wired or wireless connectivity solutions depending on deployment needs. It is an important aspect for local network deployments as, for example, factory or building owners can use the most suitable connectivity to data cloud or local servers. Additionally, integration to 3GPP networks can be enabled based on non-3GPP access interworking architectures enabling cellular operators to integrate local DECT-2020 networks as part of their infrastructure.

The mesh network topology supports direct communication between RDs as well as group communication without a dedicated D2D operation mode, enabling different use cases not possible with the star topology, such as lighting control. The strong benefit of mesh topology is that range and coverage can be extended by adding more nodes to the network. Therefore, communication with low-quality radio links canbe avoided with the benefit of removing the single points of failure within the network.

*B. Radio Interface Architecture*

The radio interface architecture of DECT-2020 mMTC Standard is shown in Fig. 2. It consists of two layers, the physical layer and the medium access control (MAC) layer.

The new DECT-2020 physical layer, designed to support 17 different frequency bands, is suitable for operating in frequency bands below 6 GHz. The employed RF technologies for the physical layer are time division duplex mode and cyclic prefix orthogonal frequency division multiplexing (CP-OFDM) with multiple access schemes utilizing both time division multiple access and frequency division multiple access. The base subcarrier spacing (SCS) is 27 kHz obtained with a scaling factor of 1. The SCS scaling supports the power of two scaling with 2, 4, 8 as other possible values. The base fast Fourier transform (FFT) size is 64 obtained with a bandwidth scaling factor of 1. Other possible scaling values are 2, 4, 8, 12, 16 resulting in the minimum channel bandwidth of 1.728 MHz with a maximum of 221.184 MHz.

The DECT-2020 NR utilizes a physical layer packet structure depicted in Fig. 2. It consists of two separate parts: the synchronization training field (STF) and the data field (DF). The STF part is used for the per-packet synchronization of the receiver and to indicate the demodulation reference signal (DRS) configuration used to transmit the physical layer packet. The DF part of the physical layer packet offers two different channels: the physical control channel (PCC) and the physical data channel (PDC). The DF part carries DRS for up to 8 antenna ports for channel estimation and equalization at the receiver. The PCC uses 98 subcarriers, where the physical control field bits, set by the MAC layer, are turbo encoded as specified in [11, Part 3].

The physical control field has either 40 or 80 information bits referred to as Type 1 and Type 2, respectively. The receiver decodes blindly whether the physical layer packet uses Type 1 or Type 2 content allowing the reception of both types without any foreknowledge. The first type is optimized for beacon transmission with a lower coding rate to provide improved coverage for PCC detection of beacon transmission. The second type is used when PDC carries higher-layer data such as unicast or multicast data. The Type 2 is designed to support advanced physical layer functionalities, as discussed further. For operation in license-exempt bands with multiple overlapping networks, the PCC carries a short Network ID and the 16-bit receiver and transmitter IDs.

The physical layer specified in [11, Part 3] is also responsible for performing the following functions: (i) error detection on the physical channel, (ii) forward error correction encoding/decoding of the physical channel, (iii) hybrid automatic repeat request (HARQ) soft-combining, (iv) rate matching of the coded physical channel data to physical channels, (v) mapping of the coded physical channel data onto physical channels, (vi) modulation and demodulation of physical channels, (vii) frequency and time synchronization, (viii) radio characteristics measurements, (ix) multiple-in multiple-out (MIMO) transmit diversity and beamforming. DECT-2020 combines the latest physical layer technologies to inbuilt mesh system architecture to future IoT solutions. Variants of these physical layer technologies have been previously available only in 3GPP cellular solutions or latest amendments of IEEE 802.11, and have never been utilized in mesh-based sensor network solutions.

*C. MAC Protocol and Coexistence*

The DECT-2020 MAC protocol, see Fig. 2, supports both the license-exempt and licensed spectrum operation. This has been taken into account in network and RD addressing as well as in radio resource management. For network addressing, the DECT-2020 uses a 32-bit network ID where 24 most significant bits (MSB) are used to identify DECT-2020 network globally and transmitted in Network and Cluster Beacon Messages. The 8 least significant bit (LSB) Network ID is selected locally and transmitted in PCC. For RD addressing, DECT-2020 uses a 32 bit Long RD ID that identifies RD uniquely in a DECT-2020 network. Furthermore, each RD has a 16 bit short RD ID, randomly selected by the RD. It is used in PCC for identifying the transmitter and receiver of a packet. Both short and long RD IDs are used during association signaling to specify the linkage between short and long RD IDs.

Following [11, Part 4] for radio resource management, RD may initiate an FT mode operation. In this mode, RD coordinates local radio resources and provides information about how other RDs may connect and communicate with it. In the FT mode, RD performs the following functions: (i) carrier and sub-slots selection, (ii) transmitting beacons to enable other RDs to identify, measure, and initiate an association with it, (iii) configuring the radio communication parameters of the connections with associated RDs, (iv) HARQ retransmissions and feedback signaling in PCC, (v) link adaptation by selecting optimum modulation and coding scheme (MCS) and transmission power. An RD in PT mode operates based on the information provided by RDs in FT mode. Note that RD can operate both in FT and PT modes simultaneously, enabling cluster-tree mesh network topology (see Fig. 1).

According to [11, Part 4], MAC supports the listen before talk (LBT) with binary exponential back-off and scheduled transmission regimes. Both random access channel (RACH) transmission and scheduled transmission use the same physical layer packet format and support HARQ. The RD in FT mode can control the number of resources allocated for both types of transmissions and assign scheduled resources for any associated RD. For RACH, RD in FT mode indicates the slots and sub-slots when RACH transmission is possible. The transmitting RD selects the MCS of the transmission and the actual packet duration to prevent exceeding the maximum allowed resources set by the RD in FT mode. Before the actual RACH transmission, an RD performs LBT to avoid collisions. The minimum sensing time for the LBT in RACH transmission is designed to be as short as two symbols, allowing for rapid access to the medium.

For the scheduled access, the RD in FT-mode uses the MAC Resource allocation information element (IE) to assign RD-specific resources in sub-slots for operation. The transmitting RD performs a selection of MCS, MIMO mode, and Tx power. The resource duration can be assigned with sub-slot accuracy with desired repetition and validity in time. The validity of allocation can be configured as single transmission and permanent allocation, based on service needs and spectrum regulation, as the RD in FT-mode needs to restrict the validity and perform LBT before re-scheduling resources again. This functionality can be utilized for voice or audio communications and industrial applications with deterministic delay requirements. Additionally, it can be used in sensor applications, where resource allocation is performed once for a group of nodes.

To enable coexistence and interference avoidance, DECT-2020 supports two types of received signal strength indicator (RSSI) measurements as defined in [11, Part 2]. The RSSI-1 is generic for interfering signals of any kind and is used in LBT. In RSSI-2, the measurement is done for any DECT-2020 signal, where PCC is decoded. As PCC contains both a short network ID and a 16-bit transmitter and receiver IDs, an RD can associate the obtained measurement report with the correct DECT-2020 network (own or other) rather than considering this interference like any arbitrary interference. This functionality is utilized in the DECT core frequency band, 1880–1900 MHz in Europe, and when operating at a spectrum with local or shared license schemes. For operation in DECT core bands, the channel access rules have additional features to detect whether slots are used by legacy DECT.

IV. PERFORMANCE INSIGHTS AND COMPARISON

In this section, we provide (i) assessment of the critical performance metrics including PLR, 99 latency percentile, and energy consumption, (ii) show the ranges of device densities, where ITU-R parameters are satisfied, and (iii) whenever possible, compare DECT-2020 with other technologies.

A. Simulation Environment

The system model has been implemented using system-level simulation tool WINTERsim extended to capture the specifics of the DECT-2020 technology. WINTERsim is a discrete event system-level (NS3-like) Python3 based simulation tool. For the DECT-2020, we used node abstraction as the basic class for all types of network devices, which contains all the node parameters (e.g., position, mobility model, antenna parameters, power) and serves as a container for L2 interfaces, routing protocol implementations, flow tables, signaling engine, application handlers, traffic generators, physical layer interfaces, etc. To simulate such a high-density deployment, we implemented specific physical layer abstraction, which stores all the bidirectional pathloss states between the RDs, having reception signal strengths higher than the noise level. During the simulation, all the components can gather component-specific and global statistics on packet deliveries, delays, power consumption, etc.

TABLE I. System Parameters for the Conducted Performance Evaluation

| Parameter | Value |
|---|---|
| **Baseline evaluation configuration parameters** | |
| System Architecture | Mesh, 19 gateways, 57 sinks |
| Carrier frequency | 1900 MHz |
| Number of channels | 1 and 3 |
| Channel bandwidth | 1.728 MHz |
| Channel model, Sink-Node | Urban macro |
| Channel model, Node-Node | Urban street canyon |
| Base station (BS) antenna height | 5 m and 25 m |
| Total Tx Power per TRxP in BS/sink | 7 and 17 dBm |
| Node power class | 7 dBm |
| **Additional parameters for system-level simulation** | |
| Inter-site distance (sink) distance | 500 m |
| Node density | 0.1-24 million devices per km |
| Number of node/sink antenna elements | 1 |
| Inter-site interference modeling | Explicitly modeled. |
| BS noise figure | 7 dB |
| End node noise figure | 7 dB |
| BS/sink antenna element gain | 0 dBi |
| RD antenna element gain | 0 dBi |
| Thermal noise level | −174 dBm/Hz |
| Traffic model | 1 message/2 hours/node |
| Application data message size | 32 bytes |
| **PHY Design** | |
| MCS | Quadrature phase-shift keying (QPSK) ¾ |
| Transmission of (negative) acknowledgements ACK/NACK | Single slot using QPSK ¾ |
| Maximal number or retransmissions | 3 |
| End node antenna height | 1.5 m |
| Frame size | 10 slots (10 ms) |
| Slot size | 10 symbols (416 us) |
| Transport block size | 456 bits |
| SCS | 27 kHz |
| Modulation scheme | CP-OFDM |

The simulation procedure starts with node placement and path loss calculations, which are done according to 3GPP guidelines [12]. The simulations have been performed according to urban macro-mMTC test Configuration A from ITU-R guidelines [8]. Particularly, it consists of standard 19 cells deployment with an inter-site distance of 500 m. RDs in TBSs are assumed to be equipped with omnidirectional antennas. As neither 3GPP nor ITU-R describes the procedure for calculating D2D links specifically, we use different models for different types of links: for $RD_{FT}$-to-any we use the Urban macro model for $RD_{FT,PT}$-$RD_{PT}$ and $RD_{FT,PT}$-$RD_{FT,PT}$ we used Urban Micro model, if both RDs are indoor and closer than 25m we used Indoor Office model, for each indoor-to-outdoor penetration we also added random building penetration loss according to 3GPP model. The percentage of indoor devices is equal to 80. The simulation procedure consists of two periods: the "warm-up" period when all the RDs establish a tree topology using different biases and the actual simulation. The default system parameters are provided in Table I.

To obtain the metrics of interest, we employ the following procedure. For each set of parameters, we ran simulations for 25 hours of system time. The steady-state period has been detected by utilizing the moving average statistics, and the data have been collected during the steady-state period. To remove residual correlations, we have utilized the batch means strategy. Due to the large size of statistical samples, only the point estimates are shown.

With Release 16, 3GPP has introduced modifications to LTE-M and NB-IoT to reduce (Radio Resource Control) RRC signaling and physical layer control channel overhead to improve spectral efficiency and device power consumption. The 3GPP NR Industrial IoT (IIoT) in Release 16 primarily focuses on Augmented reality (AR) and Virtual Reality (VR) and new use cases like factory automation to differentiate from low power use cases further [13]. Performance of these improvements introduced to LTE-M and NB-IoT and NR IIoT solution would be interesting separate studies. However, we chose to use 3GPP Release 15 as a benchmark, as it was submitted to ITU-R as 5G proposal by 3GPP. More specifically, we consider the LTE-M technology [14] and a single-hop regime of DECT-2020 technology. To enable the comparison, we adjust the amount of emitted power such that no cell-edge users experience outages. For the multi-hop regime, we chose the radiated power value to be ten times lower than for single-hop, which results in only a third of the nodes having direct connectivity to a sink.

*B. Numerical Assessment*

We start by analyzing the PLR as a function of node density for several operational regimes demonstrated in Fig. 3. First, by comparing mesh-based multi-hop and single-hop solutions, we note that the mesh-based solution shows better results. Particularly, for the single-channel regime, the density of nodes satisfying the one percent PLR requirement [5] increases more than twice from seven million for the single-hop solution to the maximum of 16 million for DECT-2020. However, the available resources increase from one to three channels does not lead to substantial performance improvement. The rationale is that each Base Station (BS) is allowed to use only one channel. Particularly, with respect to requirements of one percent PLR, the density of end nodes increases by at most one million for all considered configurations. Note that these results are in-line with [15] that studies the system in similar conditions.

Now, we introduce the bias parameter for the multi-hop operation: zero-bias means operation at the receiver sensitivity level, while, e.g., 3 dB bias means operating 3 dB above the receiver sensitivity. The usage of higher bias values forces nodes to choose next-hop neighbors that are closer to them. In practice, this feature can be implemented by setting the minQuality threshold in beacons as specified in [11, Part 4]. Observing Fig. 3, one may notice that the usage of bias improves the system capacity in the multi-hop regime. For example, for the single-channel regime, by increasing the bias from 3 dB to 20 dB, we improve the device density from 13 to approximately 17 million. Finally, note that the gain caused by bias is similar for one- and three-channel cases.

Another critical performance measure is latency. To this aim, Fig. 4 shows the 99th latency percentile. As one may observe, single-hop communication leads to better delay performance for the low density of end nodes. Raising the bias from 3 dB to 20 dB increases the delay as the data are transmitted by utilizing more hops on average. At the same time, the 99th latency percentile is not severely affected by the number of channels utilized, as for higher device density, starting from approximately half a million, all curves coincide, leading to similar performance. However, even for high device densities of around 10 million, where PLR is still within the requirement of one percent, see Fig. 3, the latency of successfully delivered packets is within 10 seconds, even for the single-hop regime. The comparison between the DECT-2020 mMTC solution and LTE-M technology is of special interest. By utilizing the LTE-M performance evaluation results [14] for a single 1.4 MHz channel, we observe that DECT-2020 outperforms LTE-M in terms of delay: 3–7 ms for DECT-2020 operational modes and slightly over than 1 s for LTE-M for the node density of half a million devices. As LTE-M uses repetitions resulting in high delay even in low load conditions, DECT-2020 avoids such operation, and it eventually turns out into reduced latency.

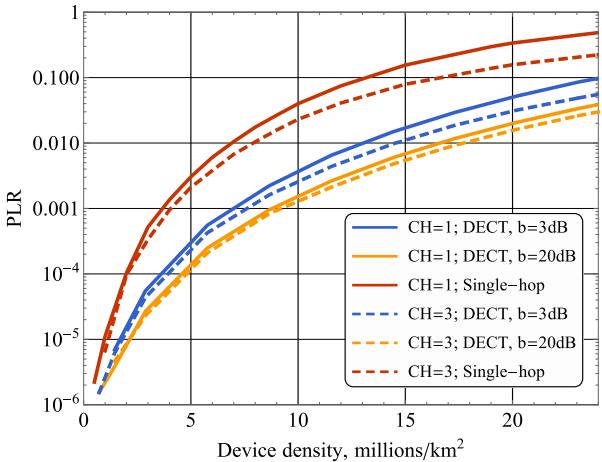

Fig. 3. Packet loss rate as a function of node density.

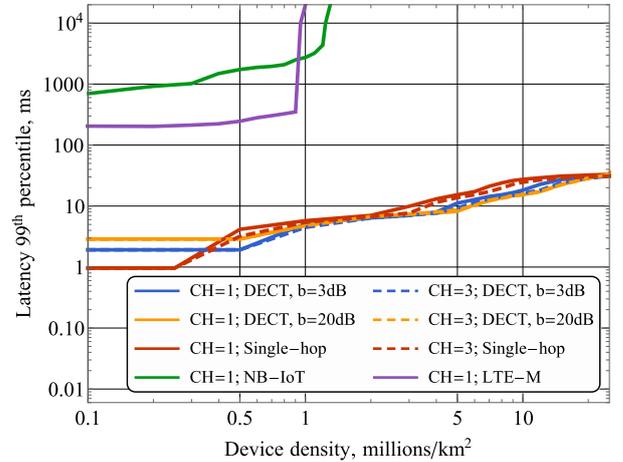

Fig. 4. 99th latency percentile as a function of node density.

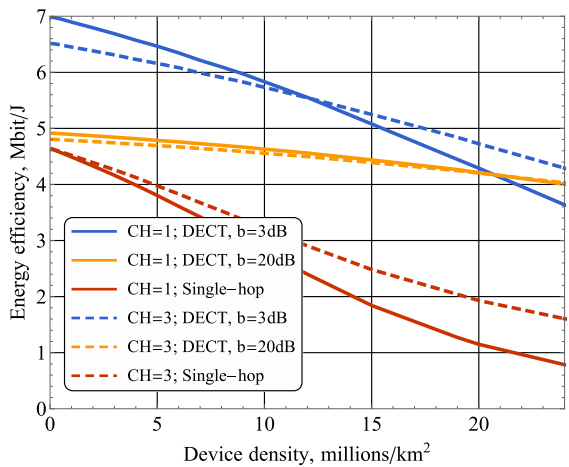

Fig. 5. Energy efficiency as a function of node density.

A critical metric responsible for an end node lifetime is energy efficiency, presented in Fig. 5 in Mbit/J. Firstly, note that the single-hop configuration leads to much worse performance than all considered multi-hop DECT-2020 setups and device densities. This effect is explained by the fact that the amount of energy needed to reach a BS in a single hop is much higher. The gradual decrease in the energy efficiency for all considered schemes is caused by the increase in the interference level, eventually leading to more frequent link-layer retransmissions. By comparing DECT-2020 multi-hop configurations, we observe that the higher bias value leads to worse energy efficiency. The reason is that higher biases require a higher number of hops on average and thus higher energy per bit. However, as the device density increases, the mode with 20 dB bias approaches the one with 3 dB bias. The rationale is that this configuration leads to lower interference, and this gain outweighs the one associated with the decrease in the number of hops.

We note that even the single-channel regime of DECT-2020 satisfies ITU loss and delay requirements for 5G mMTC communications [5]. However, from the energy efficiency point of view, this regime is much worse compared to multi-hop mode. Furthermore, single-hop communication end-systems have to be more complex, allowing to reach high emitted power.

## V. Conclusions

In this paper, we introduced the recently standardized DECT-2020 solution for novel massive IoT applications requiring stricter delay and loss performance guarantees. Utilizing a mesh topology to reduce interference, DECT-2020 can manage these deployments thanks to dynamic channel selection, cognitive radio principles without a need for precise frequency planning. Together with the state-of-the-art radio capabilities, these features make it a very attractive option for the local area, low latency, self-hosted mMTC, and other IoT networks.

Benchmarking the performance of the DECT-2020 solution against operator-driven CIoT solutions demonstrates that the mesh-based approach allows reaching much higher node densities for the same delay and loss performance bounds, enabling wide-area networking for many different applications. From the analysis we conclude, that DECT-2020 standard is best suitable for smart IoT applications such as homes, buildings, cities, industrial automation, etc., performing tracking, metering, and control tasks.

One of the open questions is the support of extremely high mobility scenarios, e.g., enabling connectivity at highways and high-speed railroads, where mesh solutions may show sub-optimal performance. Specifically, if the routing nodes move too fast, it may negatively affect the stability of the network topology. However, if only leaf nodes are highly mobile, they can still use the uplink connection; such features as local broadcasting might also enable the downlink. To solve these issues, additional changes to the routing procedures are needed.

BIBLIOGRAPHIES

**Roman Kovalchukov** is a Doctoral Researcher at Tampere University. His research interests are in wireless communications, focusing on machine learning, queuing theory, and system-level simulations.

**Dmitri Moltchanov** is University Lecturer at Tampere University. His research interests include 5G/5G+ systems, Terahertz communications, industrial IoT applications, and mission-critical V2V/V2X systems.

**Juho Pirskanen** is a Principal System Engineer at Wirepas, Finland. His research interests are different radio communication solutions on physical and radio protocol layers for new wireless standards.

**Joonas Säe** is a Staff Scientist at Tampere University. His research interests include 5G mobile networks, IoT technologies, and UAV communications.

**Jussi Numminen** is a Head of radio strategy and IPR at Wirepas, Finland. His research interests are radio communications performance on physical and radio protocol layers for new wireless standards.

**Yevgeni Koucheryavy** is a Full Professor at Tampere University. His research interests include Terahertz communications, 5G/5G+ systems, and nanonetworks.

**Mikko Valkama** is a Full Professor and Head of the Electrical Engineering Unit at Tampere University, Finland. His research interests include radio communications, localization, and sensing, with emphasis on 5G and beyond mobile networks.